\documentclass[prl,twocolumn,nofootinbib,notitlepage,10pt,aps,reprint]{revtex4-1}
\usepackage{epsfig,amsmath,amssymb,slashed}
\usepackage[utf8]{inputenc}
\usepackage{hyperref}
\usepackage{natbib}
\usepackage{amsthm}
\usepackage{xcolor}

\newcommand{\di}{{\rm d}}
\newcommand{\D}{{\rm d}}

\def\wT{{\widehat T}}
\def\wj{{\widehat j}}

\def\wP{{\widehat P}}

\def\wrhol{{\widehat{\rho}_{\rm LE}}}

\newcommand{\tr}{{\rm tr}}  
\newcommand{\Tr}{{\rm Tr}}

\newcommand{\be}{\begin{equation}}
\newcommand{\ee}{\end{equation}}                                                                               
\newcommand{\bea}{\begin{eqnarray}}
\newcommand{\eea}{\end{eqnarray}}

\begin{document}

\title{Local polarization and isothermal local equilibrium in relativistic 
heavy ion collisions}

\author{F. Becattini}\affiliation{Universit\`a di 
 Firenze and INFN Sezione di Firenze, Via G. Sansone 1, 
	I-50019 Sesto Fiorentino (Florence), Italy} 
\author{M. Buzzegoli}\affiliation{Universit\`a di 
 Firenze and INFN Sezione di Firenze, Via G. Sansone 1, 
	I-50019 Sesto Fiorentino (Florence), Italy}
\author{G. Inghirami}\affiliation{GSI Helmholtzzentrum f\"ur 
Schwerionenforschung GmbH,\\Planckstr. 1, 64291 Darmstadt , Germany}
\author{I. Karpenko}\affiliation{Faculty of Nuclear Sciences and Physical 
Engineering, Czech Technical University in Prague,\\  B\v rehov\'a 7, 11519 Prague 1, Czech Republic}
 \author{A. Palermo}\affiliation{Universit\`a di 
 Firenze and INFN Sezione di Firenze, Via G. Sansone 1, 
	I-50019 Sesto Fiorentino (Florence), Italy}

\begin{abstract}
We show that the inclusion of a recently found additional term of the spin polarization vector 
at local equilibrium which is linear in the symmetrized gradients of the velocity field, and 
the assumption of hadron production at constant temperature restore the quantitative agreement 
between hydrodynamic model predictions and local polarization measurements in relativistic heavy 
ion collisions at $\sqrt{s_{\rm NN}} = 200$ GeV. The longitudinal component of the spin polarization 
vector turns out to be very sensitive to the temperature value, with a good fit around 155 MeV. 
The implications of this finding are discussed.
\end{abstract}

\maketitle

\textit{Introduction} -
\belowpdfbookmark{Introduction}{MarkIntro}
Spin polarization in a relativistic fluid has been observed in the Quark Gluon Plasma
(QGP) formed in relativistic heavy ion collisions \cite{star,Adam:2018ivw}. From a theory standpoint,
the quantitative tool to calculate the expected polarization has mostly been a local equilibrium 
formula relating the mean spin vector of a particle with four-momentum $p$ to the 
thermal vorticity at the leading order \cite{Becattini:2013fla}. For a spin $1/2$
particle this reads:
\be\label{spinvarpi}
 S_\varpi^\mu(p)= - \frac{1}{8m} \epsilon^{\mu\rho\sigma\tau} p_\tau 
 \frac{\int_{\Sigma} \di \Sigma \cdot p \; n_F (1 -n_F) \varpi_{\rho\sigma}}
  {\int_{\Sigma} \di \Sigma \cdot p \; n_F},
\ee
where thermal vorticity is defined as the anti-symmetric derivative of the four-temperature
field:
\be\label{thvort}
  \varpi_{\mu\nu} = -\frac{1}{2} \left( \partial_\mu \beta_\nu - \partial_\nu \beta_\mu \right).
\ee
The four-temperature vector is related to the four-velocity $u$ and the comoving temperature $T$
by $\beta^\mu=u^\mu /T$. In the \eqref{spinvarpi}, $n_F$ is the Fermi-Dirac phase-space distribution 
function: $n_F = \{\exp[\beta\cdot p - q \mu/T]+1\}^{-1}$.

The measured global spin polarization of $\Lambda$ hyperons, integrated over all momenta, 
turns out to be in quantitative agreement with the predictions of the formula~\eqref{spinvarpi},  
the thermal vorticity field being provided by hydrodynamic simulations~\cite{Karpenko:2016jyx,Xie:2017upb,Wu:2019eyi,Ivanov:2020udj,Fu:2020oxj} 
and by other models \cite{Li:2017slc,Wei:2018zfb,Vitiuk:2019rfv}. However, the predicted spin 
polarization as a function of momentum, the so-called local polarization, disagrees with the 
measurements. Particularly, the sign of the longitudinal component of the spin polarization
vector and the trend of the component perpendicular to the reaction plane as 
a function of the azimuthal angle are opposite to the model predictions \cite{Becattini:2020ngo}. 

\begin{figure}
    \centering
    \includegraphics[width=0.4\textwidth]{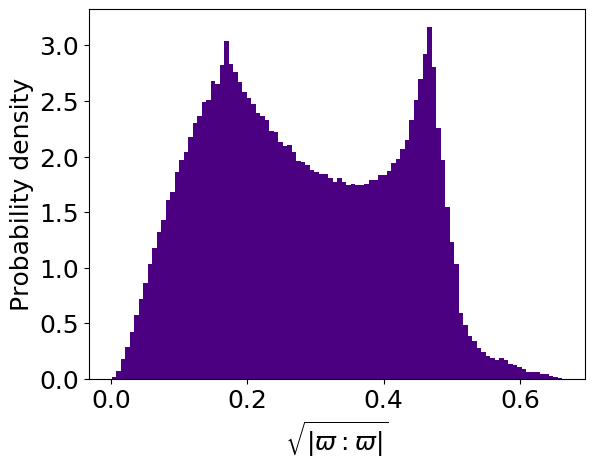}\\
     \includegraphics[width=0.4\textwidth]{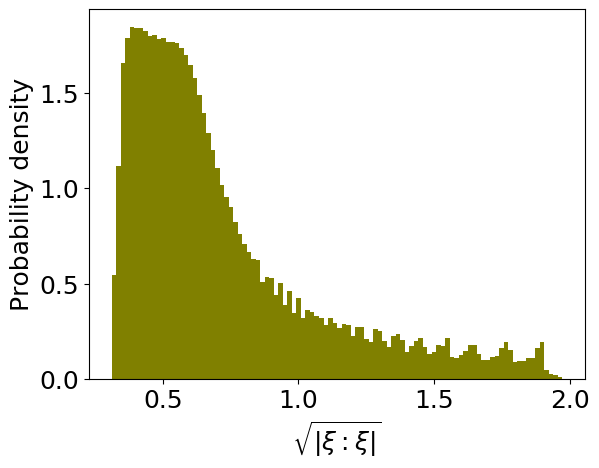}
    \caption{Distributions of the magnitudes of the thermal voriticity $\sqrt{|\varpi:\varpi|}$ (top) 
    and thermal shear $\sqrt{|\xi:\xi|}$ (bottom) at the decoupling hypersurface for a
    temperature of 165 MeV and impact parameter 9.2 fm at $\sqrt s_{\rm NN} = 200$ GeV. }
    \label{fig:derivatives}
\end{figure}

For the QGP, the formula \eqref{spinvarpi} is applicable to the final hadrons, 
provided that $\Sigma$ is identified with the hadronization 3D hypersurface or, more
rigorously, the hypersurface where the system ceases to be a fluid at local thermodynamic
equilibrium. The failure of \eqref{spinvarpi} in reproducing local polarization 
stimulated much work in the field. While it has become clear that hadronic decays 
cannot be responsible for the discrepancies \cite{Becattini:2019ntv,Xia:2019fjf}, 
investigations have been undertaken on the impact of dissipative corrections 
\cite{Hattori:2019lfp,Weickgenannt:2020aaf,Bhadury:2020puc,Bhadury:2020cop,Shi:2020htn}, of hadronic 
interactions \cite{Csernai:2018yok,Nogueira-Santos:2020aky}, on kinetic equilibration 
\cite{Gao:2019znl,Zhang:2019xya,Liu:2019krs,Li:2019qkf,Kapusta:2019ktm} and on the possible 
role of the spin tensor and an associated spin potential 
\cite{Florkowski:2017ruc,Florkowski:2019qdp,Florkowski:2019voj,Weickgenannt:2020aaf,
Gallegos:2021bzp}. One may wonder whether the inclusion of quadratic and higher order 
terms in thermal vorticity would fix the discrepancies. However, thermal vorticity
 - which is adimensional in natural units - 
is definitely less than 1 over the decoupling hypersurface (see fig. \ref{fig:derivatives}) 
and subleading corrections are not expected to cure the problem. 

\textit{Spin polarization from thermal shear} -
\belowpdfbookmark{Shear}{MarkShear}
Lately, it has been observed that, at the linear order in the gradients, there is an unexpected 
additional contribution to spin polarization vector at local equilibrium \cite{1852607,Liu:2021uhn}.
In the derivation of ref.~\cite{1852607}, for a spin $1/2$ particle, this reads:
\begin{equation}\label{spinxi}
S_\xi^\mu(p)= -\frac{1}{4m} \epsilon^{\mu\rho\sigma\tau} \frac{p_\tau p^\lambda}{\varepsilon}
 \frac{\int_{\Sigma} \di \Sigma \cdot p \; n_F (1 -n_F) 
 \hat{t}_\rho\xi_{\sigma\lambda}}{\int_{\Sigma} \di \Sigma \cdot p \; n_F}
\end{equation}
where $\varepsilon=\sqrt{m^2 + {\bf p}^2}$, $\hat{t}$ is the time direction in the QGP or
center-of-mass frame, and $\xi$ is the symmetric derivative of the four-temperature, defined as
{\em thermal shear} tensor:
\be
  \xi_{\mu\nu} = \frac{1}{2} \left( \partial_\mu \beta_\nu + \partial_\nu \beta_\mu \right) .
\ee
This additional term is also a purely local equilibrium term, i.e. non-dissipative, and
provides a contribution to the spin which is comparable to \eqref{spinvarpi} 
(see fig.~\ref{fig:derivatives}). It is obtained \cite{1852607} by expanding the thermodynamic 
field $\beta$ in the Local thermodynamic Equilibrium (LE) density operator:
\be\label{rhole}
  \wrhol = \frac{1}{Z_{\rm LE}} \exp \left[ - \int_\Sigma \di \Sigma_\mu \; 
  \left( \wT^{\mu\nu} \beta_\nu - \wj^\mu \zeta \right) \right]  
\ee
from the point $x$ where the Wigner function $W^+(x,p)$, entering the general expression 
of the spin polarization vector:
\begin{equation}\label{spingeneral}
S^\mu(p)=\frac{1}{2}\frac{\int_\Sigma \D\Sigma \cdot p\,\tr\left[\gamma^\mu\gamma^5 W^+(x,p)\right]}
    {\int_\Sigma \D\Sigma\cdot  p\, \tr\left[W^+(x,p)\right]},
\end{equation}
must be evaluated. In the eq.~\eqref{rhole}, $\wT$ is the symmetrized Belinfante
stress-energy tensor operator, $\wj$ is a conserved current and $\zeta T$ its associated
chemical potential.
It is worth dwelling into some mathematical details behind the formulae \eqref{spinvarpi} 
and \eqref{spinxi}. When evaluating the mean value of a local quantum operator 
(such as the Wigner function) at some space-time point $x$ with the LE density operator~\eqref{rhole},
i.e. $O(x) = \Tr (\wrhol \widehat O(x))$, in the hydrodynamic limit of slowly varying 
$\beta$ and $\zeta$ fields, one can obtain a good approximation by Taylor expanding the fields 
at the same point $x$ and replacing a truncated expansion at some order in the exponent 
of \eqref{rhole} as well as in the partition function $Z$ \cite{Becattini:2014yxa,Becattini:2020sww}. 
For instance, at the first order:
$$
 \beta_\nu(y) \simeq \beta_\nu(x) + \partial_\lambda \beta_\nu(x) (y-x)^\lambda
$$
and replacing into the \eqref{rhole}, with $\zeta=0$ which is a good approximation for the 
purpose of this work:
\begin{align}\label{rholeappr}
& \wrhol \simeq \frac{1}{Z_{\rm LE}} \exp \left[ - \beta_\nu(x)  \wP^\nu + \right. \\ \nonumber 
 & - \left. \partial_\lambda \beta_\nu(x) \int_\Sigma \di \Sigma_\mu(y) (y-x)^\lambda \wT^{\mu\nu}(y) 
 \right] , \\ \nonumber 
& = \frac{1}{Z_{\rm LE}} \exp \left[ - \beta(x) \cdot \wP - \left( \frac{1}{T} \partial_\lambda u_\nu(x) 
 + u_\nu \partial_\lambda (1/T) \right) \right. \\ \nonumber 
& \left. \times \int_\Sigma \di \Sigma_\mu(y) (y-x)^\lambda \wT^{\mu\nu}(y) \right]
\end{align}
In the above equation, $\wP^\nu$ is the four-momentum operator, $\beta_\nu(x) \wP^\nu$ 
is the usual global thermodynamic equilibrium exponent with constant four-temperature 
equal to $\beta(x)$ and the second term in the exponent is the leading gradient correction;
also, the contributions from temperature gradient and velocity gradient have been 
split for later use.
Retaining the zeroth order of the above expansion corresponds to the so-called perfect fluid 
approximation; going to higher orders implies including non-dissipative quantum corrections 
in the local equilibrium calculations. Since the gradient of $\beta$ supposedly gives rise 
to a small correction to the leading term $\beta_\nu(x) \wP^\nu$, one can handle it as a 
perturbation and apply linear response theory to obtain an approximation of the mean value 
which is linear in the gradient. 
Starting from the general formula \eqref{spingeneral} and splitting $\partial \beta$ in the
eq.~\eqref{rholeappr} into an anti-symmetric and a symmetric part eventually yield the two 
terms \eqref{spinvarpi} and \eqref{spinxi} respectively~\cite{1852607}. It is also worth
pointing out that the latter term stems from a correlator with a non-conserved integral 
operator \cite{1852607}, which explains the appearance of one particular vector $\hat t$ in
the formula \eqref{spinxi}; this vector can be interpreted as a sort of mean normal vector
perpendicular to the hypersurface $\Sigma$.

\textit{Gradient expansion for relativistic nuclear collisions at very high energy} -
\belowpdfbookmark{Gradient expansion}{MarkGradient}
In the formulae \eqref{spinvarpi}, \eqref{spinxi} and \eqref{spingeneral}, the hypersurface
$\Sigma$ should be the decoupling hypersurface in order to be applicable to the quasi-free 
hadronic effective fields and, at the same time, to be close to local thermodynamic equilibrium.
The question may arise whether the formulae above are the actual best approximations at the 
leading order in the gradients for the conditions of a relativistic nuclear collision
at very high energy. In fact, since the decoupling expectedly occurs at constant temperature, 
(temperature is the only effective intensive variable as chemical potentials are 
negligible), the approximation can be improved. Indeed, if $\Sigma$ is an isothermal 
hypersurface, the constant temperature $T$ in $\beta^\mu=(1/T)u^\mu$ can be 
taken out of the integral of the LE density operator:
\be\label{rhole2}
  \wrhol = \frac{1}{Z_{\rm LE}}
  \exp \left[ - \frac{1}{T} \int_\Sigma \di \Sigma_\mu \; \wT^{\mu\nu} u_\nu \right] 
\ee
and one can expand in a Taylor series only the four-velocity $u$, which
is not constant over $\Sigma$. Hence, the approximation \eqref{rholeappr} 
is replaced with:
\begin{align}\label{rholeappr2}
 \wrhol \simeq
 & \frac{1}{Z_{\rm LE}} \exp \left[ - \beta_\nu(x)  \wP^\nu + \right. \\ \nonumber 
 & - \left. \frac{1}{T} \partial_\lambda u_\nu(x) \int_\Sigma \di \Sigma_\mu(y) 
 (y-x)^\lambda \wT^{\mu\nu}(y) \right]. 
\end{align}
Comparing the equations, it can be seen that the term proportional to the gradient 
of temperature in \eqref{rholeappr} disappeared in the \eqref{rholeappr2}. 
However, it should be emphasized that the temperature gradient $\partial T$ itself
does not vanish and it is indeed perpendicular to the hypersurface $T={\rm const}$. 
Consequently, the term proportional to the gradient of temperature 
in the expansion \eqref{rholeappr} is non-vanishing even if the hypersurface $\Sigma$
is $T={\rm const}$ and it eventually contributes to both the spin polarization vector 
expressions in eqs.~\eqref{spinvarpi} and \eqref{spinxi}. In fact, the inclusion 
of such a term obtained by expanding in full space-time a function which is constant 
over the hypersurface, makes the first-order approximation of the actual LE \eqref{rhole} 
a worse one, as it introduces a term which would eventually be cancelled in the 
full Taylor series. In other words, it is not necessary, neither is it a good 
approximation, to expand in space-time a function which is constant over some 
sub-manifold of space-time (like a 3D hypersurface) if one has to integrate over 
that sub-manifold. In conclusion, the 
approximation \eqref{rholeappr2} is more accurate than the \eqref{rholeappr} for the density 
operator \eqref{rhole} if $\Sigma$ is a $T={\rm const}$ hypersurface. With the 
(approximated) density operator \eqref{rholeappr2}, it is straightforward to obtain 
the spin polarization vector in the linear response theory; comparing the 
\eqref{rholeappr} with the \eqref{rholeappr2}, all we have to do is to make the 
effective replacement:
$$
  \partial \beta \rightarrow \frac{1}{T_{\rm dec}} \partial u
$$
where $T_{\rm dec}$ is the constant decoupling temperature, in both the equation 
\eqref{spinvarpi}and \eqref{spinxi}. Particularly, the spin polarization vector 
of an emitted spin $1/2$ baryon becomes:
\begin{align} \label{spinreal}
 & S_{\rm ILE}^\mu(p) = \\ \nonumber
 &- \epsilon^{\mu\rho\sigma\tau} p_\tau 
  \frac{\int_{\Sigma} \di \Sigma \cdot p \; n_F (1 -n_F) 
  \left[ \omega_{\rho\sigma} + 2\, \hat t_\rho \frac{p^\lambda}{\varepsilon} 
\Xi_{\lambda\sigma} \right]}
  { 8m T_{\rm dec} \int_{\Sigma} \di \Sigma \cdot p \; n_F}
\end{align}
where ILE stands for {\em isothermal local equilibrium},
$$
 \omega_{\rho\sigma} = \frac{1}{2} \left(\partial_\sigma u_\rho - \partial_\rho u_\sigma \right)
$$
is the kinematic vorticity and:
$$
\Xi_{\rho\sigma} =  \frac{1}{2} \left(\partial_\sigma u_\rho + \partial_\rho u_\sigma \right)
$$
is the kinematic shear tensor. It should be stressed that the equation \eqref{spinreal} 
is not equal to the sum of $S^\mu_\varpi(p)$ and $S^\mu_\xi(p)$ integrated over 
the $T={\rm const}$ hypersurface, what is confirmed by numerical computation (see fig.~\ref{fig:Pi2D-2}).
Therefore, the equation \eqref{spinreal} is the best approximation of the spin polarization
vector of a spin $1/2$ baryon, at local equilibrium and at linear order in the 
gradients of the thermodynamic fields for an isothermal decoupling hypersurface.
This equation upgrades the original \eqref{spinvarpi} and we are going to show 
that it is able to restore the agreement between the hydrodynamic model and the data.

\textit{Analysis of Au-Au collisions at $\sqrt{s_{\rm NN}} = 200$ GeV} -
\belowpdfbookmark{Analysis}{MarkAnalysis}
To compare the predictions of the hydrodynamic model with typical initial conditions 
with the polarization data, we have used two different 3+1 D viscous hydrodynamic codes
in the Israel-Stewart formulation: vHLLE \cite{Karpenko:2013wva} and ECHO-QGP 
\cite{DelZanna:2013eua,Inghirami:2016iru}. The parameters defining the initial 
conditions have been set to reproduce charged particle multiplicity distribution 
in pseudo-rapidity as well their elliptic flow and 
directed flow in Au-Au collisions at $\sqrt{s_{\rm NN}} = 200$ GeV. 

In order to match the experimental conditions of the local polarization measurements of 
$\Lambda$ hyperons \cite{Adam:2019srw}, we set the same centrality range in our hydrodynamic 
simulations, corresponding to 20-60\% central Au-Au collisions. vHLLE simulations have been 
initialized with averaged entropy density profile from the Monte Carlo Glauber model, generated 
by GLISSANDO v.2.702 code \cite{Rybczynski:2013yba}; ECHO-QGP has been initialized with 
optical Glauber initial conditions by using the same method as in ref.~\cite{Becattini:2015ska}, 
with a fixed impact parameter $b$ set to 9.2 fm. 

\begin{figure}
    \centering
    \includegraphics[width=0.5\textwidth]{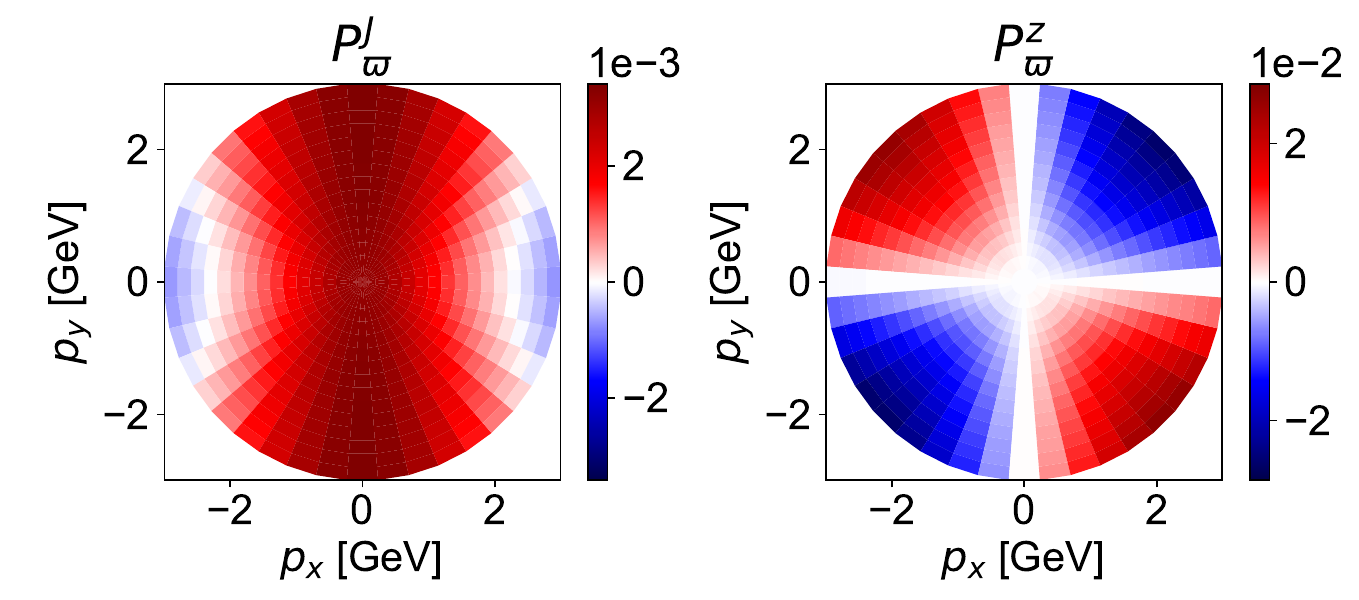}\\
    \includegraphics[width=0.5\textwidth]{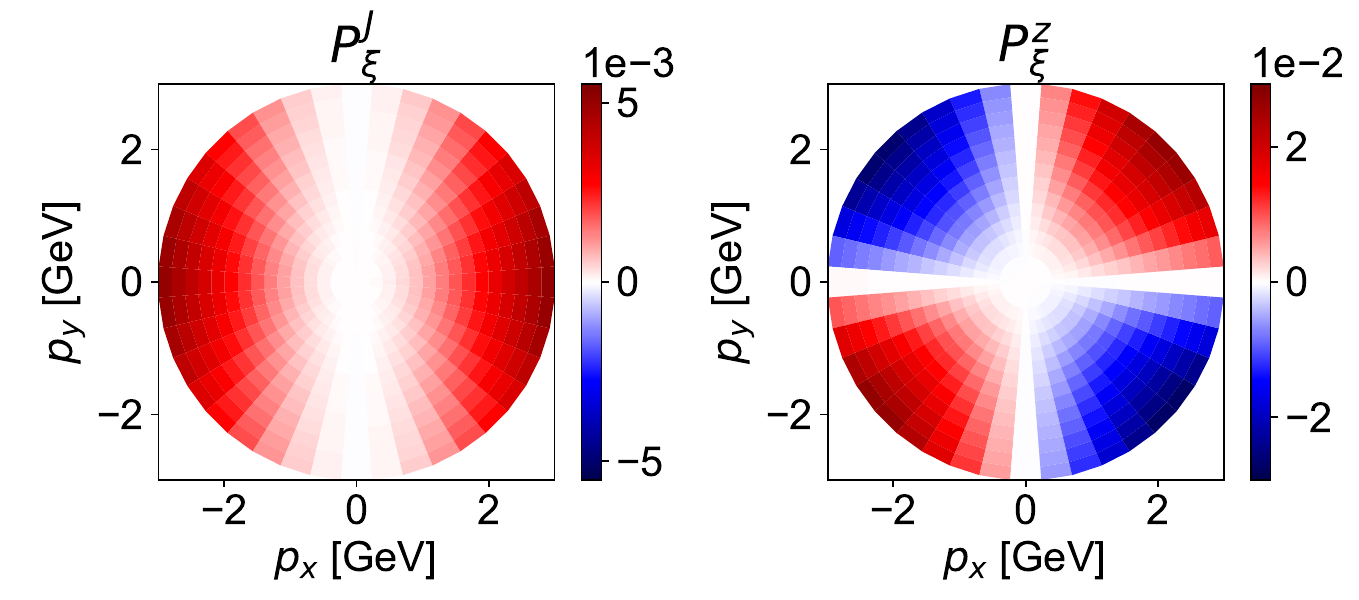}
    \caption{$\Lambda$ polarization components at mid-rapidity as a function of its transverse 
    momentum $(p_x,p_y)$, computed with vHLLE for 20-60\% Au-Au collisions at 
    $\sqrt{s_{\rm NN}}=200$~GeV. Upper panel: polarization induced by thermal vorticity 
    $\varpi$, lower panel: polarization induced by thermal shear $\xi$.}
    \label{fig:Pi2D-1}
\end{figure}
\begin{figure}
    \centering
    \includegraphics[width=0.5\textwidth]{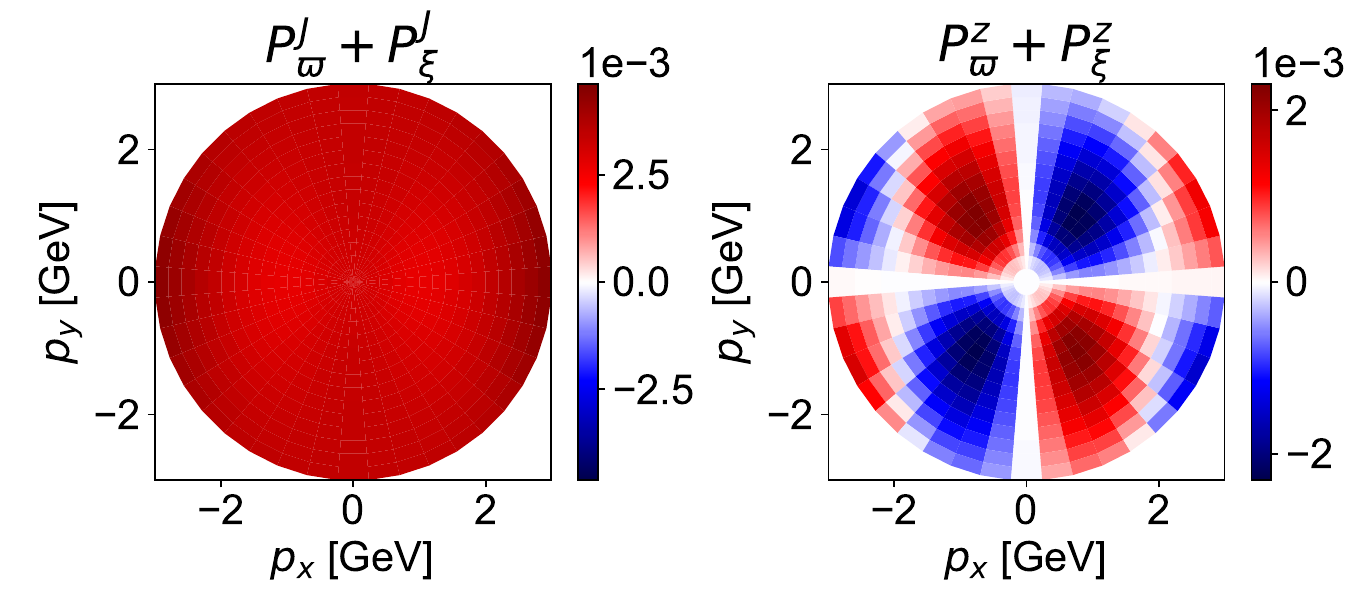}\\
    \includegraphics[width=0.5\textwidth]{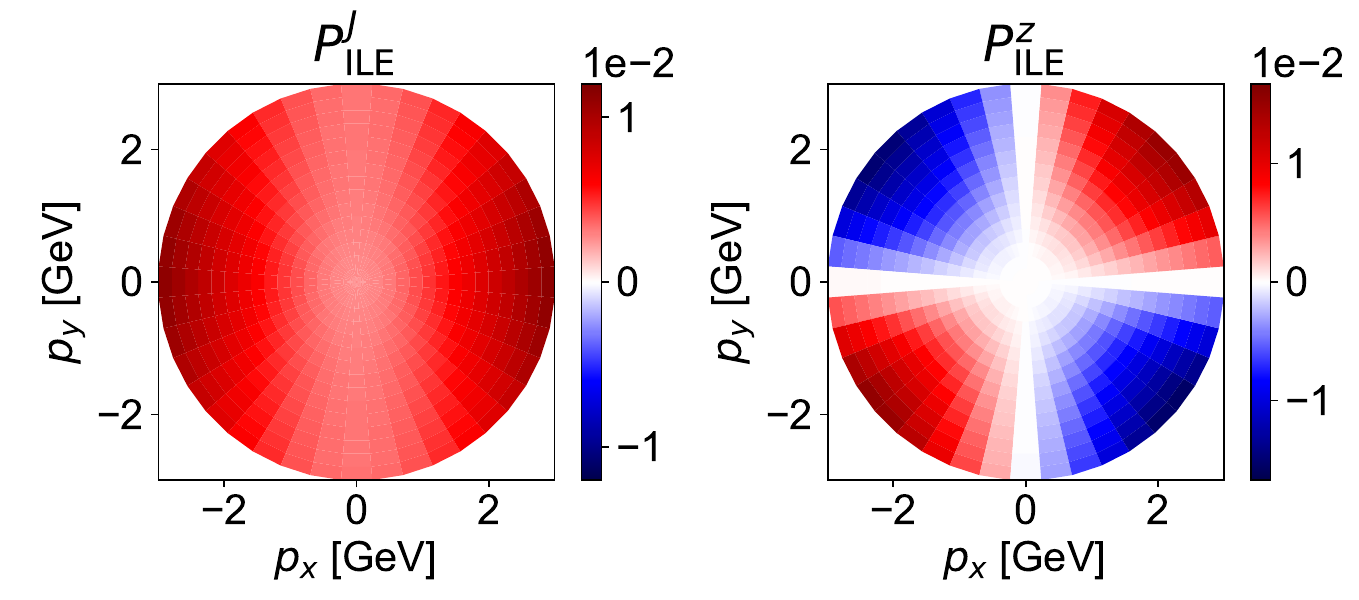}
    \caption{Same as Fig.~\ref{fig:Pi2D-1}, with the upper panels showing the sum 
    of $S^\mu_\varpi$ and $S^\mu_\xi$ from equations \eqref{spinvarpi} and 
    \eqref{spinxi}; the lower panels show the predictions of eq.\ \eqref{spinreal}.}
    \label{fig:Pi2D-2}
\end{figure}
\begin{figure}
    \centering
    \includegraphics[width=0.5\textwidth]{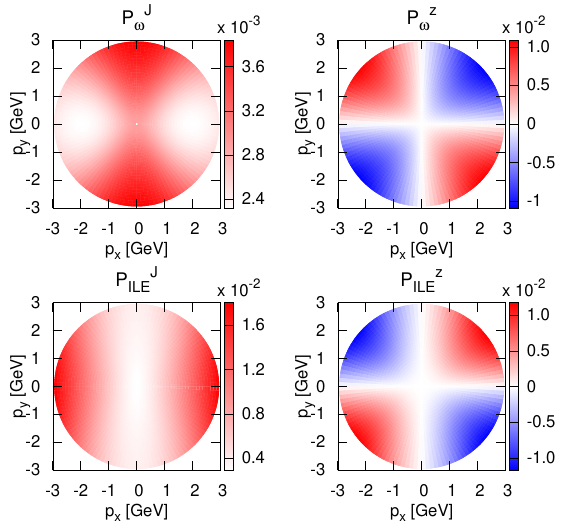}
    \caption{$\Lambda$ polarization components at mid-rapidity as a function of its transverse 
    momentum $(p_x,p_y)$, computed with ECHO-QGP. Upper panel: contribution from the first term
    in equation~\eqref{spinreal} induced by $\omega/T$. Lower panel: full prediction of
    equation~\eqref{spinreal}.}
    \label{fig:echo-pola}
\end{figure}
In figure~\ref{fig:Pi2D-1} we show the components of the rest-frame polarization vector 
${\bf P} = 2 {\bf S_*}$ along the angular momentum $P_J$ and along the beam direction 
$P_z$ (for the description of the QGP conventional reference frame, see \cite{Becattini:2017gcx})
as a function of the transverse momentum of the $\Lambda$ hyperon for rapidity $y=0$, from vHLLE 
calculation. The upper panels show the predictions of the formula \eqref{spinvarpi}, and the lower panels
the predictions of the new term \eqref{spinxi}, at a decoupling temperature $T_{\rm dec} = 165$
MeV. The two contributions are comparable in magnitude and, most importantly, the new term 
provides a local polarization in qualitative agreement with the data \cite{Adam:2019srw,Niida:2018hfw}, 
both for the $P_J$ and the $P_z$ components, and in agreement with a very recent analysis 
\cite{Fu:2021pok} of the thermal shear contribution. 
The two terms are added up and the result shown in the upper panels of the figure~\ref{fig:Pi2D-2}.
It can be seen that, although the model predictions are somewhat closer to the experimental 
findings, there is still a consistent discrepancy: a basically uniform $P_J$ \cite{Niida:2018hfw} 
and still the wrong sign of $P_z$ \cite{Adam:2019srw}. Finally, by using the formula \eqref{spinreal}, 
based on isothermal local equilibrium, we obtain polarization distributions, shown in the lower 
panels of fig.~\ref{fig:Pi2D-2}, which are in an agreement with the measurements, with the right 
sign of $P_z$ and the qualitatively correct $P_J(\phi)$ dependence. These findings are confirmed 
by a corresponding analysis made with the ECHO-QGP code and shown in fig.~\ref{fig:echo-pola}.
\begin{figure}
    \centering
    \includegraphics[width=0.5\textwidth]{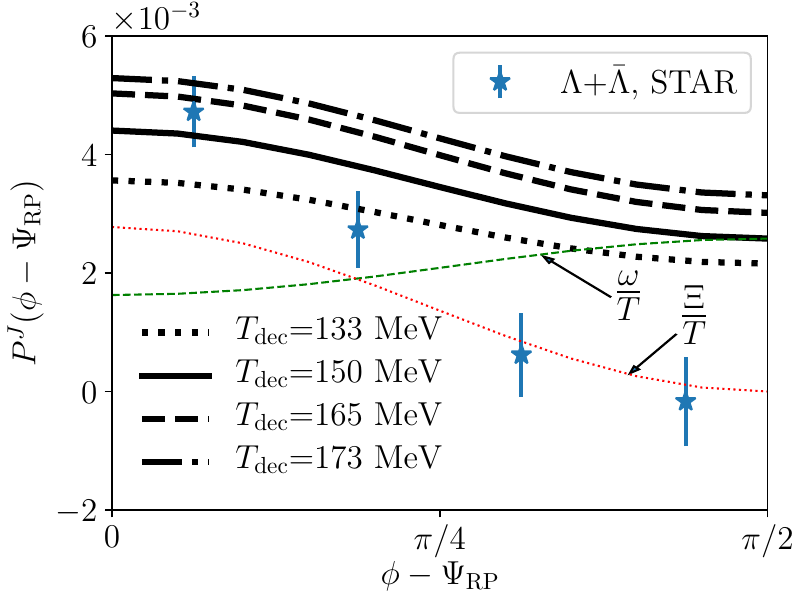}
    \caption{$\Lambda$ polarization component along the global angular momentum, as a function 
    of the azimuthal angle $\phi$, computed with vHLLE for 20-60\% Au-Au collisions at 
 $\sqrt{s_{\rm NN}}=200$~GeV. Experimental data points are taken from \cite{Niida:2018hfw}.}
    \label{fig:Pj-noTgrad}
\end{figure}
\begin{figure}
    \centering
    \includegraphics[width=0.5\textwidth]{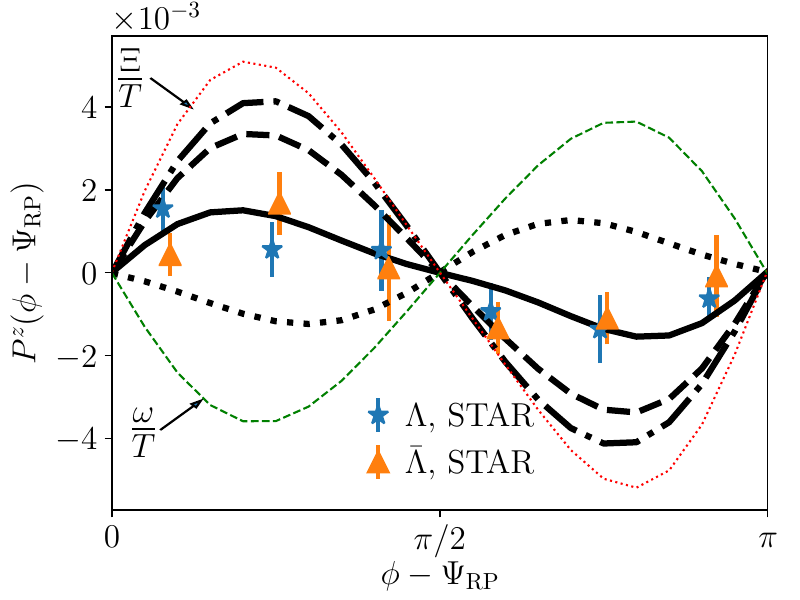}
    \caption{$\Lambda$ polarization component along the beam direction, as a function 
    of the azimuthal angle $\phi$, computed with vHLLE for 20-60\% Au-Au collisions 
at $\sqrt{s_{\rm NN}}=200$~GeV. Experimental data points are taken from \cite{Adam:2019srw} 
and conversion from $\langle\cos\theta_p^*\rangle$ to $P_H$ is performed using 
$\alpha_{\rm H}=0.732$ \cite{PDG:2020}. Error bars represent the sum of statistical 
and systematic uncertainties. Line styles correspond to different decoupling temperatures 
as in Fig.~\ref{fig:Pj-noTgrad}.}
    \label{fig:Pz-noTgrad}
\end{figure}

Finally, we have compared the data with the predictions of the eq.~\eqref{spinreal} 
at four different decoupling temperatures in figs.~\ref{fig:Pj-noTgrad} 
and \ref{fig:Pz-noTgrad} by integrating the $p_T$ spectrum of the $\Lambda$ in the same range as 
in the data, that is 0.5--6 GeV. It can be seen that the longitudinal component $P_z$ is very 
sensitive to the decoupling temperature, and it is in very good agreement with the data, for 
$T_{\rm dec}$ value around 150-160 MeV; for temperatures below around 135 MeV, the sign of the 
longitudinal polarization flips.
The $P_J$ component is now predicted to have a maximal value on the reaction plane, in agreement
with the data, however with a milder descent as a function of the azimuthal angle; also, it is 
less sensitive to $T_{\rm dec}$. We also note that the global polarization 
resulting from the integration of $P_J$ is still in a reasonably good agreement 
with previous calculations. Also shown, in both figures, are the contributions from the 
kinematic vorticity $\omega$ (thin dashed line) and the kinematic shear $\Xi$ (thin smaller 
dashed line), at the decoupling temperature of 150 MeV. It can be seen in figure \ref{fig:Pz-noTgrad} 
that the latter is crucial to flip the sign of $P_z$ and restore the agreement with the data, 
while the vorticity term alone would give the wrong sign, as already remarked in 
ref.~\cite{Wu:2019eyi}.

\textit{Discussion, conclusions and outlook} -
\belowpdfbookmark{Discussion}{MarkDiscussion}
The recently found additional shear term and the realization of the constancy 
of $T_{\rm dec}$ are the two key ingredients to reproduce the local polarization 
and the $P_J$ and $P_z$ patterns.
This finding is thus a striking confirmation of the local equilibrium picture or,
in perhaps more suggestive words, the quasi-ideal fluid paradigm of the QGP, even 
in the spin sector. Dissipative corrections to spin polarization may play 
a role, but they appear not to be decisive. The standard hydrodynamic picture with the 
initial conditions obtained by fitting radial spectra, elliptic and directed flow, 
works very well for the local polarization too.
Another strong indication from this finding is that, at very high energy, the QGP hadronizes 
in space-time at constant $T_{\rm dec}$ to a more accurate level than one could have 
imagined. Indeed, its sensitivity to the {\em gradients} of the thermodynamic fields, makes
spin the ideal probe to investigate the space-time details of hadron formation. Furthermore, 
as we have shown, the longitudinal spin polarization turns out to be very sensitive to the 
decoupling temperature, the causes of which deserve to be studied in
detail. Looking ahead to future investigations, it is certainly important to compare the predictions
of the formula \eqref{spinreal} as a function of transverse momentum and rapidity besides 
azimuthal angle. At lower energy, where the chemical potentials are relevant, one can expect a decoupling
hypersurface different from the simple $T={\rm const}$, and this will require a reconsideration 
of the \eqref{spinreal} in order to obtain accurate predictions. 

\begin{acknowledgments}
\textit{Acknowledgments} -
G.\ Inghirami acknowledges funding by the Deutsche Forschungsgemeinschaft (DFG, German 
Research Foundation) – Project number 315477589 – TRR 211. M.B. is supported by the Florence 
University fellowship {\em Effetti quantistici nei fluidi relativistici}.
I.K.\ acknowledges support by the project Centre of Advanced Applied Sciences with 
number CZ.02.1.01/0.0/0.0/16-019/0000778, which is co-financed by the European Union, 
and support by the Ministry of Education, Youth and Sports of the Czech Republic under 
grant ``International Mobility of Researchers – MSCA IF IV at CTU in Prague'' 
No.\ CZ.02.2.69/0.0/0.0/20\_079/0017983. Computational resources were supplied by 
the project ``e-Infrastruktura CZ'' (e-INFRA LM2018140) provided within the program 
Projects of Large Research, Development and Innovations Infrastructures.

\end{acknowledgments}

\bibliographystyle{apsrev4-1}


\end{document}